\documentclass[12pt]{article}
\title{Non-commutative Gross-Neveu model at large $N$}

\author{Emil T. Akhmedov\footnote{Permanent Address:
Institute of Theoretical and Experimental Physics,
B. Cheremushkinskaya 25, 117259 Moscow, Russia},
Philip DeBoer and  Gordon W. Semenoff\\~~\\
Department of Physics and Astronomy, \\
University of British Columbia, \\
Vancouver, British Columbia, Canada V6T 1Z1.}

\begin{document}           

\maketitle                 

\abstract{The non-commutative $O(N)$ Gross-Neveu model is solved in
the large $N$ limit in two and three space-time dimensions.  The
commutative version of the two dimensional model is a renormalizable
quantum field theory, both in a coupling constant expansion and an
expansion in $1/N$.  The non-commutative version has a renormalizable
coupling constant expansion where ultraviolet divergences can be
removed by adjusting counterterms to each order.  On the other hand,
in a previous work \cite{Akhmedov:2000uz} we showed that the
non-commutative theory is not renormalizable in the large $N$
expansion.  This is argued to be due to a combined effect of
asymptotic freedom and the ultraviolet/infrared mixing that occurs in
a non-commutative field theory.  In the present paper we will
elaborate on this result and extend it to study the large $N$ limit of
the three dimensional Gross-Neveu model.  We shall see that the large
$N$ limit of the three dimensional theory is also trivial when the
ultraviolet cutoff is removed.

\newpage

\noindent

\section{Introduction and Summary}

In a previous paper \cite{Akhmedov:2000uz} we discussed the issue of
renormalizability of the non-commutative $O(N)$-symmetric Gross-Neveu
model in the leading order of a large $N$ expansion.  The model is a
massless two dimensional field theory of fermions with a four-fermion
interaction.  It has a dimensionless coupling constant and the
commutative version is renormalizable.  The coupling constant has an
asymptotically free renormalization group flow. As a result, the
interaction is large at small momenta.  The strong coupling dynamics
in the infrared regime leads to spontaneous breaking of chiral
symmetry by dynamical generation of a fermion
mass~\cite{Gross:1974jv}.

The non-commutative version of this model also has a formally
renormalizable perturbation expansion.  Non-commutativity improves the
ultraviolet limit of the theory slightly in that, order by order in an
expansion in the coupling constant, the Feynman diagrams which have
ultraviolet divergences are a subset of those which diverged in the
commutative theory.  It is at least plausible that the ultraviolet
divergences can be removed by adjusting counterterms for vertices
which have the same form as the vertices in the tree level action
~\cite{Chepelev:2000tt} \cite{Chepelev:2000hm}.  The quantities whose
dependence on the ultraviolet cutoff is milder in the non-commutative
theory than they were in the commutative theory are more singular at
small momenta.  This is called ultraviolet/infrared (UV/IR) mixing.
It raises the possibility already noted in
\cite{Minwalla:1999px,VanRaamsdonk:2000rr} that, even if the theory is
renormalizable in the conventional sense, infrared divergences could
upset its consistency.

In our article \cite{Akhmedov:2000uz}, we addressed this question by
examining the large $N$ expansion of the non-commutative theory.  This
expansion sums a class of Feynman diagrams to all orders.  The
dimensionless expansion parameter is $1/N$.  We found that, in the
large $N$ expansion, because of the singular infrared structure of
certain diagrams, the ultraviolet cutoff could not be removed from the
effective interactions.  Forcibly taking the ultraviolet cutoff to
infinity made the effective four-fermion interaction zero.  In that
limit, the non-commutative $O(N)$ Gross-Neveu model is a trivial,
noninteracting field theory.

Our results could be criticized on two grounds.  First, though it does
not seem relevant to the issue of ultraviolet singularities, being in
two dimensions, the model is necessarily space-time non-commutative.
When time is not commutative, the action of the non-commutative field
theory contains infinite numbers of time derivatives.  This makes a
Hamiltonian interpretation of the theory difficult and it would be
expected to lead to difficulties with unitarity.  Though this has
little to do with ultraviolet divergences, it could be argued that
this will not be a sensible theory in any case because of the
unitarity problem.

Secondly, the two dimensional Gross-Neveu model necessarily has
spontaneous symmetry breaking and dynamical mass generation.  It seems
likely that the condensate which breaks the symmetry is a constant,
carrying zero momentum.  This condensate couples to the zero momentum
limit of correlation functions.  Because of UV/IR mixing, the
correlation functions are singular at zero momentum and the
ultraviolet cutoff re-appears as infrared divergences and cannot be
subtracted by counterterms.  If the condensate were not a constant it
would not be necessary to use the zero external momentum limit of
correlation functions to make the effective action.  In that case, the
problem with divergences at low momenta could be circumvented.  This
possibility is difficult to analyze since, thus far, no candidates for
a non-constant condensate have been found.

Both of these criticisms can be at least partially addressed by
considering the large $N$ limit of the non-commutative three
dimensional Gross-Neveu model.  In three dimensions, the commutative
model has a dimensional coupling constant and is not renormalizable in a 
conventional perturbation theory which expands order-by order in the
coupling.  However, it is renormalizable in the $1/N$
expansion
\cite{Parisi:1974nz,Gross:1975vu,Rosenstein:1989pt,Semenoff:1992hr,
Semenoff:1989dm}.  One can examine whether the non-commutative theory
is also renormalizable in the same expansion.

Being three dimensional, it is possible to have space-space
non-commutativity only, leaving time commutative.  There are some
general arguments that this should be a unitary
theory\cite{Gomis:2000zz}.

Moreover, the commutative three dimensional theory has a second order
phase transition between a symmetry breaking ordered phase
with dynamically generated fermion mass and a massless, symmetric
phase.  If the non-commutative theory also has a symmetric phase, it
can be studied in a context where generation of a constant condensate
is not an issue.  A question which we shall address in
this paper is whether the large $N$ expansion of the 
non-commutative theory can be renormalizable in that phase.  

The second order phase transition in the commutative three dimensional
theory occurs at an infrared stable fixed point of the renormalization
group flow of the four-fermion coupling.  Existence of the fixed point
is a result of the large anomalous dimension of the four-Fermi
operator.  At the tree level it is an irrelevant dimension four
operator.  In the large $N$ expansion it behaves like a marginally
relevant dimension three operator which flows to a fixed point where
the coupling is of order 1.  It is interesting to examine the
difference between the behavior of the non-commutative version of this
model and the two dimensional model where the dynamical mass
generation is the result of the absence of an infrared fixed point.

The great deal of attention that non-commutative field theories have
received recently is motivated by the fact that they arise as the low
energy limits of string theories with background antisymmetric tensor
fields \cite{Ho:1997jr,Connes:1998cr,Chu:2000gi,Seiberg:1999vs,
Lee:2000kj,Laidlaw:2000kb}.  They retain some of the interesting
features of string theory, such as non-locality, which can then be
studied in the simpler context of the non-commutative field theory.
Since the string theories are consistent quantum mechanical theories,
the non-commutative field theories which are their zero slope limits
should also be internally consistent.  In fact, for some theories,
unitarity has been demonstrated explicitly at one-loop order
\cite{Gomis:2000zz}.  The issue of renormalizability of these theories
asks whether the limit where the string distance scale
$\sqrt{\alpha'}$ is taken as arbitrarily small produces a field theory
with non-trivial interactions.


There are many features of non-commutative field theories which
distinguish them from their commutative analogs.  One already occurs
in scalar field theory.  The non-commutativity, which was at one time
suggested as an ultraviolet regularization of quantum field theory,
affects the spectrum and interactions at low energy scales, below the
momentum scale $1/\sqrt{|\theta|}$ set by the dimensional parameter
$\theta^{\mu\nu}$ and often below the mass scales of the lightest
particles already in the model.  This has been associated with the
phenomenon of UV/IR duality familiar from the behavior of D-branes in
string theory \cite{Minwalla:1999px,VanRaamsdonk:2000rr}.

The perturbation expansion of the non-commutative theory is similar to
the commutative one \cite{Filk:1996dm}.  The main difference is that
the vertices of the non-commutative theory differ from those in the
commutative theory by momentum-dependent phases.  These phases
generally improve the convergence of loop integrals.

In non-commutative field theory, there is a fat-graph representation
of the perturbation theory in which the Feynman diagrams can be
classified according to the genus of the Riemann surface on which they
could be drawn without crossing any lines.  Diagrams can be
classified according to whether they are planar or non-planar graphs
\cite{Filk:1996dm}.  The Feynman integrands of planar graphs are as
they were in the commutative case.  The integrands of non-planar
graphs are modified by phases containing external and internal loop
momenta.  The presence of these phases improves the high-momentum
behavior of Feynman integrals.  The most dramatic effect occurs in
diagrams which are ultraviolet divergent.  Planar diagrams diverge and 
must be defined using a high momentum cutoff, $\Lambda$, as was the
case in the commutative theory.  In the non-planar diagrams the
loop integrations are affected by phases and generally converge, the
ultraviolet cutoff being replaced by an effective cutoff $\Lambda_{\rm
eff}(p)=1/ \sqrt{1/\Lambda^2+(\theta p)^2/4}$. For any momentum in the
range $p>1/(\theta\Lambda)$ , this effective cutoff has a finite
limit, $\Lambda_{\rm eff}\sim 1/2|\theta p|$, as
$\Lambda\rightarrow\infty$.

For example, at one loop order in 4-dimensional $\phi^4$-theory, the
radiative correction to the scalar self-energy in the commutative
version is a quadratically divergent constant coming from a tadpole
diagram.  In the non-commutative theory, there are two contributions,
coming from a planar and a non-planar tadpole.  The planar graph is
again a quadratically divergent constant, as it was in the commutative
case.  The non-planar graph, even thought it is a tadpole, turns out
to be a function of external momentum.  At small momentum,
\cite{Minwalla:1999px,VanRaamsdonk:2000rr}
\begin{equation}
\Gamma^{(2)}(p)= \frac{g^2}{48\pi^2}\left(
\Lambda^2-m^2\ln\frac{\Lambda^2}{m^2}\right)+
\frac{g^2}{96\pi^2}\left( \Lambda_{\rm
eff}^2(p)-m^2\ln\frac{\Lambda_{\rm eff}^2(p)}{m^2}\right)+ \ldots
\label{twopt}
\end{equation}
Here, $m$ is the scalar field mass and $g^2$ is the dimensionless
$\phi^4$ coupling constant.  The second term has a pole at very low
momenta.  It also has a logarithmic cut singularity at small momenta.
These have been argued to arise from new degrees of freedom with
exotic dispersion relations or perhaps propagating in higher
dimensions \cite{VanRaamsdonk:2000rr}.  They have also been argued to
lead to exotic translation non-invariant ``striped'' phases of scalar
field theory \cite{Gubser:2000cd}.

Another place where ultraviolet divergences occur is in the
renormalization of dimensionless coupling constants.  The leading
corrections to the coupling constant in 4-dimensional $\phi^4$-theory
are logarithmically divergent.  The small momentum limit of the
4-point function was computed in \cite{Minwalla:1999px} as
\begin{eqnarray}
\Gamma^{(4)}(p,q,r,s)=g^2-\frac{g^2}{2\cdot 2^5\pi^2}\left\{
2\ln\frac{\Lambda^2}{m^2}+\ln\frac{1}{m^2(\theta p)^2}+\ln\frac{1}{m^2(\theta q)^2}
+
\right.
\nonumber \\
\left. +\ln\frac{1}{m^2(\theta r)^2}
+\ln\frac{1}{m^2(\theta s)^2}
+\ln\frac{1}{m^2(\theta(q+r) )^2}
+\right.
\nonumber \\
\left.
+\ln\frac{1}{m^2(\theta(q+s))^2}
+\ln\frac{1}{m^2(\theta(r+s))^2} \right\}+\ldots
\label{fourpt}
\end{eqnarray}
($p+q+r+s=0$) The first contribution is from planar diagrams and is
similar to that in the commutative theory, with slightly different
coefficient.  The other contributions are from non-planar diagrams and
they depend explicitly on the parameter $\theta$.

What is remarkable about (\ref{fourpt}) is that, in spite of the mass
gap of the bare scalar field, the effective coupling constant is
logarithmically singular at small momentum.  This logarithmic
singularity is similar to the one which occurs in the same function at
large momentum (with $m^2(\theta p)^2$ replaced by $p^2/\Lambda^2$).
It becomes large and ruins the perturbation expansion in both the
regimes of very large and very small momentum transfers.

This behavior is radically different from that of commutative field
theory where the running of coupling constants at low momenta is
cutoff by mass scales and is frozen at the mass scale of the lightest
matter particles which participate in the interaction.  For example,
in quantum electrodynamics which, like $\phi^4$-theory is infrared
free, the coupling constant runs at energies much larger than the
electron mass, but as the energy is lowered, it freezes at the value
$e^2/4\pi\sim 1/137$ and is the same at all lower energy scales.  In a
non-commutative theory, it appears that masses do not cutoff the
running of coupling constants.  We emphasize that this is a
non-perturbative issue, which occurs in addition to the perturbative
renormalizability of non-commutative field theories which has recently
been examined in detail
~\cite{Chepelev:2000tt,Chepelev:2000hm,Sheikh}.

This unusual running of the coupling at low momenta is a mirror of the
running of the coupling at high momenta which necessarily occurs when
the momentum scale is larger than any dimensional couplings or masses
but still less than the UV cutoff.  The UV/IR mixing in
non-commutative field theory seems to mirror that running with a
similar low energy behavior, regardless of the existence of masses for
the particles.

It is interesting to compare the Gross-Neveu model with the
four-dimensional $O(N)$ vector model with action 

$$ S=\int d^4x\left\{
\frac{1}{2}\partial\phi^i\cdot\partial\phi^i+
\frac{\lambda}{8N}(\phi^i\phi^i)^2\right\}
$$ 
In this model the infrared rather than asymptotically free running of
the effective coupling constant can lead to a ground state which
breaks translation invariance, but is otherwise apparently consistent
interacting theory when the cutoff is large.  The large $N$ expansion
is solved by introducing an auxiliary field $\sigma$,
$$ S=\int d^4x\left\{
\frac{1}{2}\partial\phi^i\cdot\partial\phi^i+\frac{i}{2}\sigma\phi^i\phi^i
+\frac{N}{2\lambda}\sigma^2\right\} $$ 
Then, in the leading order in large $N$, the small momentum limit of
the quadratic term in the effective action for $\sigma$
$$ \frac{N}{2}
\int d^4p \sigma^*(p) \left(
\frac{1}{\lambda}+c\ln\frac{\Lambda^2}{\mu^2} +c'\ln\frac{\Lambda_{\rm
eff}^2(p)}{\mu^2}+\ldots\right)\sigma(p) 
$$
where $\sigma(p)$ is the Fourier transform of $\sigma(x)$.  Here $c$
and $c'$ are positive constants.  They are positive because the beta
function of 4-dimensional scalar field theory is positive.  It is
infrared free.  The first logarithm on the right-hand-side comes from
a planar diagram, the second from a non-planar diagram. Only the small
momentum limit is presented.  The contribution of the planar diagram would
go to zero at large momentum.
The inverse of the bare $\phi^4$-coupling constant $\lambda$ must be
adjusted to cancel the UV singularity.  If we tune $\lambda$ to absorb
the singularity and put the cutoff to infinity, we obtain an
expression like 
$$ \frac{N}{2}\int d^4p \sigma^*(p)
\left(\frac{1}{\tilde\lambda} +c'\ln\frac{1}{2|\theta
p|\mu^2}+\ldots\right)\sigma(p) $$ 
To find a phase transition, we lower $1/\tilde\lambda$ to negative
values.  Eventually when $1/\tilde\lambda$ is negative enough,
$\sigma$ will have a tachyonic mode.  This should first occur for
modes at some finite momentum and there will be a condensate of
$\sigma(p)$ where $p$ is non-zero and $<\sigma(x)>$ is not a constant.
This is a different mechanism for breaking of translation symmetry
from the one which was found in ~\cite{Gubser:2000cd}.

Now, what happens in the Gross-Neveu model?  Let us focus on the
two dimensional case.  There, the auxiliary scalar field has the
quadratic term in its effective action
$$
\frac{N}{2}\int d^2p \sigma^*(p) 
\left( \frac{1}{\lambda}-g\ln\frac{\Lambda^2}{\mu^2}
-g'\ln\frac{\Lambda_{\rm eff}^2(p)}{\mu^2}+\ldots\right)\sigma(p)
$$
where $g$ and $g'$ are positive.  Note that the coefficients of 
the logarithms are
negative.  This is a result of the fact that the beta function is
negative and the theory is asymptotically free.  Again, we can add
a counterterm to cancel the UV singularity of the first logarithm, to
get
$$
\frac{N}{2}
\int d^2p \sigma^*(p) \left(  \frac {1}{\tilde\lambda}
-g'\ln\frac{\Lambda_{\rm eff}(p)}{\mu^2} \right)\sigma(p)
$$
Now, the infinite $\Lambda$ limit gives a finite expression, but there
is a dramatic difference from the infrared free case, the term is
always going to be negative for small enough momenta and always has a
minimum at zero momentum.  Then, at the minimum, the quadratic term in the
effective action for $\sigma$ 
diverges.  If we add a counterterm to make it finite at zero momentum,
then it would have a divergence at finite momentum.  This is what
leads to triviality.  The three dimensional case has similar behavior,
even in the phase where there should be no condensate. The details are
in the remainder of this paper.
 
\subsection{Notation}

On either non-commutative Euclidean space, $R^d$, or Minkowski space,
$M^d$, the coordinates have the algebra
\begin{equation}
 \left[ x^\mu, x^\nu \right]
= -i\theta^{\mu\nu}
\label{coords}
\end{equation}
where $\theta^{\mu\nu}$ is an antisymmetric matrix.  Being anti-symmetric,
it must have even rank.

Equivalent to imposing the algebra (\ref{coords}) on the coordinates,
non-commutativity can also be implemented by replacing the usual
product for multiplying functions by the associative, non-commutative
and non-local $*$-product,

\begin{equation}
f(x)*g(x)\equiv \lim_{y\rightarrow x}
\exp\left( -\frac{i}{2}\theta^{\mu\nu}\frac{\partial}
{\partial x^\mu}\frac{\partial}{\partial y^\nu}\right) f(x)g(y)
\label{noncom}
\end{equation}

Under an integral, we can always use integration by parts
to remove one of the $*$-products and replace it with an ordinary
product,
\begin{equation}
\int f_1*f_2*\ldots *f_{n-1}*f_n=\int f_1(f_2*\ldots *f_{n-1}*f_{n})
\end{equation}
For this reason, 
the multiplication of functions under an integral has cyclic symmetry
\begin{equation}
\int f_1*f_2*\ldots*f_{n-1}*f_n=\int f_n*f_1*f_2*\ldots*f_{n-1}
\end{equation}

Note that in two dimensions non-commutativity does not break Lorenz
invariance, $\theta_{\mu\nu} = \theta \cdot \epsilon_{\mu\nu}$.  Also,
in 1+1-dimensional Minkowski space, it would not be possible to set $\theta^{0i}$ to
zero, unless one set the entire matrix to zero so time is necessarily non-commutative
if the space is.

In three dimensions we can always take the space coordinates as
non-commuting and leave the time commutative by choosing $\theta^{0i}=0$.  In this
paper we will assume that this is the case.

\vskip 0.2 cm

\noindent

\section{The non-commutative O(N) Gross-Neveu model}

In this section we will fix the notation and define the models which
we will examine in some detail in the following Sections.  We review
some of the techniques for solving a vector model in the large $N$
limit.  The Euclidean action of the non-commutative Gross-Neveu model
is

\begin{equation}
S[\psi]=-\int d^Dx \left\{ \frac{1}{2}\sum_{j=1}^N
\bar\psi^j\gamma\cdot\partial\psi^j
+\frac{\lambda}{8N}
\sum_{ij=1}^N\bar\psi^i*\psi^i*\bar\psi^j*\psi^j\right\}
\label{gn}
\end{equation}
where $D$ is the dimension of space-time, which we assume to
have Euclidean signature.

The spinors $\psi^i$ are Majorana fermions.  They obey the constraint
$\psi= C\psi^*$ where $C$ is the charge conjugation matrix.  We use
Majorana, rather than Dirac fermions in order to generate non-planar
diagrams at the leading order in large $N$.  In two dimensions, the
Majorana spinors have two components and there is a choice of Dirac
matrices for which the spinor is real.  In three dimensions, we could take
them as having either two or four components and, for concreteness, we
will take them to have two components.  The distinction between these
two choices is not important for the arguments of this paper.

In three dimensions, with two-component spinors, the condensate that
we discuss will break space-time parity.

In two dimensions, the kinetic term in (\ref{gn}) has $O(N)_L\times
O(N)_R$ chiral symmetry.  The interaction term breaks this to
$O(N)\times Z_2$ where the $O(N)$ is a diagonal subgroup of
$O(N)_L\times O(N)_R$ and the $Z_2$ is a discrete chiral symmetry.
Under this residual chiral transformation, $\psi\rightarrow
\gamma^5\psi$ with $\gamma^5=i\gamma^1\gamma^2$.  This symmetry
forbids fermion mass and, for the fermions to become massive, the
$Z_2$ symmetry must be spontaneously broken.  The condensate
$\left<\bar\psi\psi\right>$ is an order parameter for this symmetry
breaking.

In three dimensions, with 2-component spinors, the kinetic and
interaction term have $O(N)$ symmetry.  The condensate
$<\bar\psi\psi>$ and accompanying fermion mass generation that we are
interested in studying breaks parity.

All products in (\ref{gn}) are $*$ products, as defined in
(\ref{noncom}).  When we set total derivative terms in the action
(\ref{gn}) to zero, in each term, because we can integrate by parts,
one of the $*$-products is always equal to an ordinary product.  For
this reason only the ordinary product occurs in quadratic terms.  The
quartic term can be written as $\int(\bar\psi*\psi)^2$.  We shall use
this fact when we introduce an auxiliary field.

In the non-commutative theory, there is another interaction vertex
of the form
$$
\int dx~\sum_{i,j=1}^N\bar\psi^i*\psi^j*\bar\psi^i*\psi^j
$$
which one could add to the action density.  Because of the
non-commutativity of the $*$-product it is not equivalent to the one
that is already in (\ref{gn}).  We have chosen not to add this vertex
because it is technically more complicated to deal with than the one
in (\ref{gn}). This vertex should not alter solvability of the model
in the large $N$ limit. However, it seems that there is no local
master field when it is present and it is likely that the ground state
has broken translation invariance. We shall not examine this issue
further in this paper.  Here, we note that, at least in the leading
order in large $N$, we shall find that the model (\ref{gn}) seems to
be consistent without this additional vertex - as a vertex of that
form does not seem to be generated dynamically.

\subsection{The effective action}

It is useful consider the generating functional for correlation
functions of the spinor field,
\begin{equation}
Z[\eta]=\int [d\psi] \exp\left(-S[\psi]+\int d^Dx~
\bar\eta(x)\psi(x)\right)
\label{generating}
\end{equation}
where $\eta(x)$ is an anti-commuting classical Majorana spinor.
The correlation functions are obtained by taking functional derivatives
of this generating functional,
\begin{equation}
\left.
\left< \psi(x_1)\ldots\psi(x_n)\right>=\frac{1}{Z[0]}
\frac{\delta}{\delta\bar\eta(x_1)}\ldots
\frac{\delta}{\delta\bar\eta(x_n)}Z[\eta] \right|_{\eta=0}
\end{equation}
As usual, connected correlation functions are generated by the logarithm
of the partition function,
\begin{equation}
W[\eta]=-\ln\left( Z[\eta]\right)
\end{equation}
From this, it is convenient to convert to a generating functional for
one-fermion irreducible correlation functions.  This is done using a
Legendre transform.  We define the 1-point function in the presence
of the source by
\begin{equation}
\psi_0(x)\equiv
< \psi(x) >=-\frac{\delta}{\delta\bar\eta(x)}W[\eta]
\end{equation}
Then, we consider the Legendre transform
\begin{equation}
\Gamma[\psi_0]= W[\eta] + \int d^Dx~\bar\eta(x)\psi_0(x)
\end{equation}
From this equation, we see that
\begin{equation}
\frac{\delta}{\delta\psi_0(x)}\Gamma[\psi_0]=-\bar\eta(x)
\label{ordp}
\end{equation}
is an equation which determines the classical expectation value of
$\psi$ induced by the source.  $\Gamma[\psi_0]$ is sometimes called
the effective action.  Higher functional derivatives of
$\Gamma[\psi_0]$ by $\psi_0$ give one-fermion irreducible correlation
functions.  The connected correlation functions can be reconstructed
from these by combinatorics and, in turn, one can reconstruct the full
correllators.

\subsection{Large N}

In order to solve the large $N$ limit of the model (\ref{gn}), it is
convenient to introduce an auxiliary field so that the action is
(where now we include the source term in the action)
\begin{equation}
S[\psi,\phi,\eta]=\int d^Dx \left\{ -\frac{1}{2}\sum_{j=1}^N
\bar\psi^j(\gamma\cdot\partial+*\phi*)\psi^j
+\frac{N}{2\lambda}\phi^2-\bar\eta\psi\right\}
\label{gn1}
\end{equation}
The original action (\ref{gn}) is recovered by doing the Gaussian
integral of $\phi$ in the partition function,
\begin{equation}
Z[\eta]=\int [d\psi d\phi]\exp\left(-S[\psi,\phi,\eta]\right)
\end{equation}

Alternatively, this is now a Gaussian integral for $\psi$ and we can
integrate out ${\psi}$ to get the non-local scalar field theory
with action

\begin{equation}
S[\phi,\eta]= -\frac{N}{2}{\rm Tr}\ln \left( \gamma\cdot\partial +
*\phi*\right) +\int\left(\frac{N}{2\lambda}\phi^2-\frac{1}{2}\bar\eta
\frac{1}{\gamma\cdot\partial+*\phi*}\eta\right)
\label{fermact}
\end{equation}
where $*\phi*$ denotes multiplication using the $*$-product.  Some
care will be required in defining the fermion determinant with $*\phi*$
present.  We will discuss the details below.

In the large $N$ limit, the remaining functional integral can be
evaluated by saddle point approximation.  For this, we must find a
minimum of (\ref{fermact}) as a functional of $\phi$.  Then, the large
$N$ limit of the generating functional for connected correllators is
given by
\begin{equation}
W[\eta]=\inf_{\phi}:
 -\frac{N}{2}{\rm Tr}\ln \left( \gamma\cdot\partial +
*\phi*\right) +\int\left(\frac{N}{2\lambda}\phi^2-\frac{1}{2}\bar\eta
\frac{1}{\gamma\cdot\partial+*\phi*}\eta\right)
\label{genf}
\end{equation}
where the infimum is taken while holding $\eta$ fixed.

This can be converted to the generating functional for
one-fermion-irreducible graphs by taking the explicit Legendre
transform.  The result is the elegant expression
\begin{equation}
\Gamma[\psi_0]=\inf_{\phi}:
 -\frac{N}{2}{\rm Tr}\ln \left( \gamma\cdot\partial +
*\phi*\right) +\int\left(\frac{N}{2\lambda}\phi^2-\frac{1}{2}
\bar\psi_0\gamma\cdot\partial\psi_0+\frac{1}{2}
\phi\bar\psi_0*\psi_0\right)
\label{genf1}
\end{equation}
where now the infimum over $\phi$ must be found while holding $\psi_0$
fixed.  Then, when we find the infimum, it will depend on $\psi_0$.
Substituting it back into (\ref{genf1}) gives a functional of $\psi_0$
whose functional derivatives by $\psi_0$ are the one-fermion
irreducible correlation functions to leading order in the large $N$
expansion.  The leading contribution to the correlator with n external
Fermion legs is order $N^{1-n}$ and is readily obtained from this
expression.  It is possible to obtain higher orders in $1/N$ by finding
corrections to the saddle point approximation that we have used here.
This is a systematic procedure.

It is clear that $\eta$ and $\psi_0$ are related in a complicated way.
We will always assume that $\psi_0$ goes to zero when $\eta$ goes to
zero.  Since $\Gamma[\psi_0]$ is an even function of $\psi_0$, when
$\eta=0$, $\psi_0=0$ is always a solution of (\ref{ordp}).

We will restrict our search for infima in (\ref{genf1}) to those
which give translation invariant ground states, i.e. to where the
function $\phi$ which minimizes (\ref{genf1}) goes to a constant
when $\psi_0$ goes zero.  At this point, we do not know whether there
are translation non-invariant solutions which have smaller action than
the translation invariant solution that we find.  We will not address
this issue in this Paper. Since we will find no tachyons in the
spectrum, the solution that we consider is at least a local minimum if
not an infimum.  If $\phi=M$ is a constant, its $*$-product with other
functions reduces to an ordinary product.  Then, setting $\psi_0$ to
zero for the moment, we can readily evaluate (\ref{genf1}) with
$\phi=M$.  The function which must be minimized to find the vacuum
value of $M$ is the effective potential
\begin{equation}
V_{\rm eff}(M)=  -\frac{N}{2}\frac{1}{\rm v}
{\rm Tr}\ln \left( \gamma\cdot\partial +
M\right) +\frac{N}{2\lambda}M^2
\end{equation}
where we have
divided by the space-time volume, v.
The trace is evaluated as
\begin{equation}
V_{\rm eff}=-\frac{Nd}{4}\int \frac{d^Dp}{(2\pi)^D}
\ln\left( p^2+M^2 \right)
+\frac{N}{2{\lambda}} M^2
\label{veff}
\end{equation}
where $d$ is the dimension of the Gamma matrices.  An ultraviolet
cutoff is needed to evaluate this integral.  After a cutoff is
introduced, it can easily be evaluated in 2 and 3 dimensions.  The
value of $M$ which minimizes this integral is the vacuum expectation
value of $\phi$ when the sources are set to zero.
The gap equation either has the solution $M=0$ or $M$ solves
the equation
\begin{equation}
\frac{1}{\lambda}=\frac{d}{2}\int\frac{ d^Dp}{(2\pi)^D}\frac{1}{p^2+M^2}
\label{gapequation}
\end{equation}
Then we must check whether the solution of (\ref{gapequation}) or
$M=0$ is the stable solution.

\subsection{Non-commutative fermion determinant}

When $\phi$ is not a constant, we must take some care in evaluating
the fermion determinant in order to correctly take into account the
$*$-products.  If $\phi$ has the form $\phi=M+\delta\phi$, where $M$ is
a constant, the determinant is defined by the expression

\begin{equation}
-\frac{1}{2}\ln{\rm Tr}\left( \gamma\cdot\partial + *\phi*\right)
\equiv \sum_{n=1}^\infty\frac{1}{n!}\int \delta\phi(x_1)
\ldots \delta\phi(x_n)\tau(x_1,\ldots,x_n) \label{two1}
\end{equation}
where
\begin{equation}
\tau(x_1,\ldots,x_n)=-\left(\frac{1}{2}\right)^n
\left< \bar\psi(x_1)*\psi(x_1)\ldots\bar\psi(x_n)
*\psi(x_n)\right>^{\rm conn.}_0  \label{two2}
\end{equation}
The expectation values on the right-hand-side of this equation
are connected correlators of free fermions with
mass $M$.  The correlators of free fermions of mass $M$ are defined by,
\begin{equation}
<\psi(x_1)\ldots\psi(x_n)>_0=\frac{
\int[d\psi]\psi(x_1)\ldots\psi(x_n)
e^{\int\frac{1}{2}\bar\psi(\gamma\cdot\partial+M)\psi}}
{\int[d\psi]e^{\int\frac{1}{2}\bar\psi(\gamma\cdot\partial+M)\psi}}
\end{equation}
From these, one should choose the connected ones to form the correlators on
the right-hand-side of (\ref{two2}).

Then, the equation which determines $\delta\phi(x)$ is
\begin{equation}
\sum_1^\infty\frac{1}{n!}\int dx_1\ldots dx_n\tau(x,x_1,
\ldots,x_n)\delta\phi(x_1)\ldots\delta\phi(x_n)+
\frac{1}{\lambda}\delta\phi(x)+\frac{1}{2N}
\bar\psi_0(x)*\psi_0(x)=0
\label{gapeqn}
\end{equation}
The solution of this equation should then be substituted back into
(\ref{genf1}) to find the generating functional for irreducible
correlators.

\subsection{Effective four-fermion coupling}

The equation (\ref{gapeqn}) which determines $\delta\phi(x)$ can be
solved iteratively.

To get the leading order, we define the Fourier
transform\footnote{Here we are distinguishing the function $\tau$ from
its Fourier transform by writing its space version with two arguments
and its Fourier transform in momentum space with one argument.  Which
of the two we mean in a particular place should always be clear from
the context.}
$$
\tau(x_1,x_2)= \int \frac{ d^2q}{(2\pi)^2}\tau(q)
e^{iq\cdot(x_1-x_2)}
$$
We shall also find it a useful short-hand to
write $\tau(x_1,x_2)$ as a differential operator
$$
\tau(x_1,x_2)= \tau(-i\partial)\delta(x_1-x_2)
$$
Then,
$$
\delta\phi(x)= -  \frac{\lambda}{1+\lambda\tau(-i\partial)}
\frac{1}{2N}\bar\psi_0(x)\psi_0(x) + \ldots
$$
where corrections are terms of second and higher order in $\bar\psi\psi$.q
Substituting this back into the effective action obtains
$$
\Gamma[\psi_0]=
-\int\left(\frac{1}{2}\bar\psi_0(\gamma\cdot\partial+M)\psi_0
+\frac{1}{8N}
\bar\psi_0*\psi_0\frac{\lambda}{1+\lambda\tau(-i\partial)}
\bar\psi_0*\psi_0+\ldots\right)
$$
The leading terms in this effective action have the same form as the
original action with the coupling constant replaced by an effective
coupling which depends on the momentum transfer $p$ and is given by
\begin{equation}
\lambda^{\rm eff}(p)=\frac{\lambda}{1+\lambda\tau(p)}
\label{eff4f}
\end{equation}

Thus, to find the effective four-fermion interaction, we must find the
two-point correlator $\tau(x_1,x_2)$.  Out of all the terms in
(\ref{gapeqn}), it is only this, leading one which is ultraviolet
divergent in either two or three spacetime dimensions.  It is therefore
the only one that will be affected by UV/IR mixing in this leading order
of the large $N$ expansion.

\subsection{Summary}

Let us review the procedure to be followed to get the effective
coupling constant.  We must first solve the gap equation
(\ref{gapequation}) for the fermion mass $M$ and decide whether this
solution, if it exists, or the trivial solution $M=0$ is stable.  The
stable solution is a global minimum of the effective potential in
(\ref{veff}).  Then we must choose the stable solution for $M$ and use
it to compute the function $\tau(p)$ using equation (\ref{two2}).
Using those two results, we find the effective coupling constant using
equation (\ref{eff4f}).

\section{Results in two dimensions}

In two dimensions, the effective potential is
\begin{equation}
V_{\rm eff}=-\frac{N}{8\pi}\left( M^2\ln\frac{\Lambda^2}{M^2}+M^2\right)
+\frac{N}{2{\lambda}} M^2
\label{veff2}
\end{equation}
This potential applies to either the commutative or the
non-commutative theory.  It must be minimized in order to find the
physical value of $M$.  In two dimensions, it always has a doubly
degenerate global minimum for non-zero $M$ which occurs when the gap
equation is solved,
\begin{equation}
\frac{1}{\lambda}=\frac{1}{4\pi}\ln\frac{\Lambda^2}{M^2}
\label{gap2}
\end{equation}
Both $M$ and $-M$ are solutions of this equation, reflecting the chiral
symmetry.

The equation (\ref{gap2}) is a statement of dimensional transmutation:
when (\ref{gap2}) is substituted for the coupling constant in a
physical expression in a renormalizable theory, the dependence of the
cutoff $\Lambda$ cancels and the parameter remaining is the mass
scale, $M$, in this case the dynamically generated fermion mass. Thus,
the bare dimensionless coupling $\lambda$ and ultraviolet cutoff
$\Lambda$ are traded for a dimensional parameter, $M$.  The theory
doesn't have a coupling constant, instead it has a mass scale and is
weakly coupled for processes with momenta which are greater than $M$
and strongly coupled when the momentum is less than $M$.

We will see this explicitly in the effective coupling constant for the
commutative theory, there $\Lambda$ and $\lambda$ disappear from the
effective coupling constant once the gap equation is solved.  They are
replaced by the mass scale, $M$.

Although we will not use it in the following, we could introduce a
renormalized coupling constant so that, substitution into
(\ref{veff2}) and (\ref{gap2}) we would remove the ultraviolet
divergences,
$$
\frac{1}{\lambda}=
\frac{1}{\lambda_{\rm ren}(\mu)}+\frac{1}{4\pi}\ln\frac{\Lambda^2}{\mu^2}
$$
Then, notice that, if we hold $\lambda_{\rm ren}(\mu)$ fixed as we put
the cutoff $\Lambda$ to infinity, the bare coupling constant $\lambda$
goes to zero.  This is another manifestation of asymptotic freedom. On
the other hand, if we hold the bare coupling $\lambda$ and the cutoff
$\Lambda$ fixed as we take the renormalization scale $\mu$ small, the
renormalized coupling $\lambda_{\rm ren}(\mu)$ increases and goes to
infinity at some small scale.  This is infrared slavery and the
infrared Landau pole.  Of course, the true running of the effective
coupling constant is cutoff by mass generation.

\subsection{Commutative model}

To see how this works, we first consider the commutative theory.  We
can do this by setting the matrix $\theta^{\mu\nu}$ to zero, which
amounts to treating all of the $*$-products in the above formulae as
ordinary products.  In this case,
\begin{equation}
\tau_{\rm c}(p)=- \frac{1}{2\pi}\left( \ln\frac{\Lambda
e^{1-\gamma}}{M} -\frac{ \sqrt{1+\frac{p^2}{4M^2}} }{\frac{p}{2M}}\ln
\left(\sqrt{ 1+\frac{p^2}{4M^2}}+\frac{p}{2M}\right) \right)
\label{commvacpol}
\end{equation}
where the subscript, c, denotes commutative, $\gamma$ is Euler's
constant.  Here and below we use the same regularization as in
\cite{Minwalla:1999px} and \cite{VanRaamsdonk:2000rr}.
Then, in this case, the effective coupling constant is
\begin{equation}
\lambda^{\rm eff}_{\rm c}(p)=\frac{2\pi}{\frac{ \sqrt{1+\frac{p^2}{4M^2}} }
{\frac{p}{2M}}\ln
\left(\sqrt{ 1+\frac{p^2}{4M^2}}+\frac{p}{2M}\right)+\gamma-1 }
\end{equation}
Here, we have used the gap equation (\ref{gap2}) to eliminate the
coupling constant.  Note that it also cancels the ultraviolet cutoff, leaving
only the mass parameter, $M$.
For large momenta, $p>>M$, the effective coupling
$$
\lambda^{\rm eff}_{\rm c}(p)\approx \frac{2\pi}{\ln(p/M)}
$$
is small, as we expected.  As we lower the momentum to small momenta,
$p<<M$, it increases and stops increasing when the momentum gets to
the scale $M$, where it freezes at
$$
\lambda^{\rm eff}_{\rm c}(p)\approx 2\pi/\gamma
$$

\subsection{Non-commutative model}

Now, let us examine the non-commutative theory.  There, $\tau(p)$
gets a contribution from both a planar and a non-planar graph.
The planar diagram contributes
$$
\tau_{\rm nc}^{\rm pl}(p)=
- \frac{1}{4\pi}\left( \ln\frac{\Lambda e^{1-\gamma}}{M}
-\frac{ \sqrt{1+\frac{p^2}{4M^2}} }{\frac{p}{2M}}
\ln\left(\sqrt{ 1+\frac{p^2}{4M^2}}+\frac{p}{2M}\right)
\right)
$$
This planar diagram is not affected by the non-commutativity.  That is
why the result is half of the commutative contribution
(\ref{commvacpol}).

There is also a non-planar diagram.  It depends on the non-commutativity
parameter in a non-trivial way,
$$
\tau_{\rm nc}^{\rm nonpl}(p)=- \frac{1}{4\pi}
K_0\left(2M
\sqrt{[\theta^2q^2/4+1/\Lambda^2]}\right) +~~~~~~~~~~~~~~~~~~~~~~~~~~~
$$
$$
+ \frac{1}{4\pi}\left(M^2 + \frac{q^2}{4}\right) \int^1_0 d
\alpha \int_0^{\infty} d\rho
\exp\left\{- \rho\left(M^2 + \alpha(1-\alpha)q^2 \right) -
 \frac{(\theta q)^2}{4\rho}\right\}.
$$
Here $K_0(z)$ is the modified Bessel function.  Note that, even though
the planar contribution is divergent, for non-zero momenta the
non-planar one has a finite limit as the ultraviolet cutoff is put to
infinity.  The correlator $\tau_{\rm nc}(p)$ in the non-commutative
theory is the sum of the above two contributions
$$
\tau_{\rm nc}(p)=\tau_{\rm nc}^{\rm pl}(p)+\tau_{\rm nc}^{\rm nonpl}(p)
$$
The effective four point coupling of the fermions with momentum
transfer $p$ is
\begin{equation}
\lambda^{\rm eff}(p)= \frac{1}{\frac{1}{\lambda} + \tau_{\rm nc}^{\rm
pl}(p) + \tau_{\rm nc}^{\rm nonpl}(p)}
\label{coupling}
\end{equation}
When we substitute the cut-off-dependent expression (\ref{gap2}) for
$1/\lambda$ into (\ref{coupling}), the UV cutoff dependence does not
cancel.  If for any momentum in the range
$p>1/\theta\Lambda$, the effective coupling $\lambda_{\rm
eff}(p)$ goes to zero as $\Lambda$ is taken to infinity.

Let us find UV behavior of (\ref{coupling}). In the limit
when $p^2 >> 4 M^2$, $p^2 >> 1/(\theta M)^2$ (we always assume that
$p^2<<\Lambda^2$ and $M^2<<\Lambda^2$) we have

\begin{equation}
\tau_1(p)\approx - \frac{1}{8\pi}\ln\frac{\Lambda^2}{p^2}
\quad {\rm and} \quad
\tau_2(p)\sim e^{-\theta M p}
\end{equation}
Thus

\begin{equation}
\lambda^{\rm eff}(p) \approx
\frac{1}{\frac{1}{\lambda} - \frac{1}{8\pi}\ln\frac{\Lambda^2}{p^2}}
=
\frac{8\pi}{\ln\frac{\Lambda^2p^2}{M^4}},
\end{equation}
where $\lambda$ is eliminated using (\ref{gap2}).
On the other hand, when $p^2<<1/(\theta^2M^2)$, $p^2 << 4 M^2$ and
$p^2 >> 1/(\theta \Lambda)^2$ we can approximate the above expressions
by
\begin{eqnarray}
\tau_{\rm nc}^{\rm pl}(p) \approx - \frac{1}{8\pi}\ln\frac{\Lambda^2}{M^2}
\quad {\rm and} \quad
\tau_{\rm nc}^{\rm nonpl}(p) \approx - \frac{1}{8\pi}
\ln\left(\theta^2p^2M^2\right)
\end{eqnarray}
Hence
\begin{eqnarray}
\lambda^{\rm eff}(p) \approx \frac{1}{\frac{1}{\lambda}
- \frac{1}{8\pi}\ln\frac{\Lambda^2}{M^2}
- \frac{1}{8\pi} \ln\left(\theta^2p^2M^2\right)} =
\frac{8\pi}{\ln\left(\Lambda^2 \theta^2 p^2\right)}.
\end{eqnarray}
For momenta above $p\sim 1/\theta\Lambda$ the last expression depends
on the cutoff and for finite, nonzero momentum it goes to zero as the
cutoff goes to infinity.

Thus, we find that, in the non-commutative theory, renormalization
does not remove the cutoff dependence of the effective four-fermion
interaction.  The interaction is suppressed by an inverse power of the
ultraviolet cutoff and goes to zero - rendering the theory trivial -
as the cutoff is put to infinity.

\subsection{What if we choose $1/\lambda$ so that the cutoff cancels?}

It is the solution of the gap equation (\ref{gap2}) which dictated the
cutoff dependence of $\lambda$.  If we choose a different cutoff
dependence for $1/\lambda$, the gap equation would not be satisfied.  This
means that the system would be unstable.  We would see this immediately in
the effective coupling constant (\ref{coupling}) - in fact the stability
condition for quadratic fluctuations of $\phi(x)$ in the action is the
positivity condition for inverse of the effective coupling.

For example, we could remove the cutoff dependence by choosing
$$
\frac{1}{\lambda}=\frac{1}{\tilde\lambda(\mu)}
+\frac{1}{8\pi}\ln\frac{\Lambda^2}{\mu^2}
$$
Then, we can take the cutoff to infinity.  We find, at very small momenta
$$
\lambda^{\rm eff}(p)\approx \frac{1}{\frac{1}{\tilde\lambda(\mu)}
+\frac{1}{8\pi}\ln(\mu^2p^2\theta^2/4)}
$$
This effective coupling exhibits an infrared Landau pole at
a small value of the momentum squared,
$$
p_L^2=\frac{4}{\theta^2\mu^2}e^{-8\pi/\tilde\lambda(\mu)}
$$
Modes of $\phi(x)$ with momenta less than this value are tachyonic,
a result of the instability caused by not using a proper solution of
the gap equation.

\subsection{A double scaling limit}

Quite interesting things happen in the double scaling limit when
$\Lambda\to\infty$ and $\theta\to 0$ so that $\Lambda\theta = C/M$
with an arbitrary constant $C$. The physical meaning of this limit is
that one ``regularizes'' the ordinary Gross-Neveu model by a
non-commutative one at the cutoff scale.  This theory is
non-commutative only at distance scales of order of and smaller than
the ultraviolet cutoff.  But in field theory, the UV/IR mixing of
large and small momentum scales means that it still has an effect.  In
this particular limit we can obtain an exact expression:
\begin{eqnarray}
\lambda^{\rm eff}(q) =
\frac{4\pi}{\frac12\ln\left(1 + C^2 q^2/M^2\right) +
\frac{ \sqrt{1+\frac{q^2}{4M^2}} }
{\frac{q}{2M}}\ln\left(\sqrt{ 1+\frac{q^2}{4M^2}}+\frac{q}{2M}\right)}.
\end{eqnarray}
The second term in the denominator has a square root cut starting from
$q=2Mi$ in the complex $q$ plane, which corresponds to a pair
production of fermions.  This is the same as what occurs in the
commutative Gross-Neveu model.  What is new and interesting is the
first term.  It has a logarithmic cut starting from $q=iM/C$. This cut
is absent in the commutative model. We speculate that it corresponds
to creation of pairs of some non-local solitons present in the
non-commutative theory, which survive this double scaling limit.
However, as yet we have not been able to find an explicit form for the
objects which are created, if indeed they are solitons.

We see that the limits $\Lambda\rightarrow\infty$ and
$\theta\rightarrow 0$ do not commute and even in the case when
non-commutativity is relevant at the cutoff scale it still modifies
the behavior of the theory at any energy scale.

\section{Results in three dimensions}

In the three dimensional Gross-Neveu model, the coupling constant
$\lambda$ has the dimension of length.  This model is therefore not
renormalizable in a perturbation theory in the coupling constant.
However, in the large $N$ expansion of the commutative model, the only
counterterms needed to cancel the divergences that arise are of the
same form as the operators which are already in the action.  Thus, by
the usual definition, the commutative theory is a renormalizable model
in the $1/N$ expansion.  The reason for its renormalizability is the
existence of a second order phase transition which occurs at a
sufficiently large value of the coupling constant.  It is near that
value of the coupling that the theory is renormalizable.  At the
second order phase transition, the four-fermion coupling has a large
anomalous dimension, so it is effectively
dimensionless\cite{Chen:1993ig}.

In three dimensions, the gap equation (\ref{gapequation}) with
a suitable cutoff is
\begin{equation}
0 = M\left[\frac{1}{\lambda} - \frac{1}{16\pi} \Lambda
\exp\left(-\frac{2M}{\Lambda}\right)\right].
\label{veff1}
\end{equation}
Unlike in two dimensions, we see that in three dimensions, which of the
two possible solutions of (\ref{veff1}) is stable, depends on the size of
the coupling constant. If
$$
\lambda< \frac{16\pi}{\Lambda}
$$
the solution is
$$
M=0
$$
In this phase, the fermions are massless and the chiral symmetry is unbroken.

If, on the other hand,
$$
\lambda > \frac{16\pi}{\Lambda}
$$
then $M$ solves the equation
\begin{equation}
\frac{1}{\lambda} = \frac{1}{16\pi} \Lambda
\exp\left(-\frac{2M}{\Lambda}\right). \label{gap1}
\end{equation}
In this phase, the fermions have mass and the chiral
symmetry is spontaneously broken.

\subsection{Commutative model}

Let us first examine the commutative model.  There are two phases which
are separated by a second order phase transition.

\subsubsection{Symmetric, massless phase}

First, consider the symmetric phase, where $\lambda<16\pi/\Lambda$.
In this case, the correlation function $\tau_{\rm c}(q)$ which
determines the effective coupling is readily computed,
\begin{equation}
\tau_{\rm c}(q)=-\frac{\Lambda}{16\pi}+\frac{ |q|}{16}
\end{equation}
and the effective coupling constant is
$$
\lambda^{eff}(q)= \frac{1}{\frac{1}{\lambda}
-\frac{\Lambda}{16\pi}+\frac{ |q|}{16} }
$$
We tune $\lambda$ to be sufficiently close to
the critical value, and of course less than the critical
value,
\begin{equation}
\frac{1}{\lambda}=\frac{\Lambda}{16\pi}+\frac{\mu^2}{16}~~,~~\mu\geq 0
\label{coupren}
\end{equation}
This is equivalent to adding a counterterm to the action.  The result is
$$
\lambda^{eff}(q)= \frac{16}{\mu^2+|q| }
$$
At the critical point, when $\mu^2=0$ this is a non-trivial conformal
field theory whose scaling properties can be computed order by order
in the $1/N$ expansion \cite{Chen:1993ig,Petkou:1996ad,Petkou:1999wd}.
At the critical point, the fermion mass operator $\bar\psi\psi$ has
conformal dimension 1, rather than its tree level value of $2$.

Note that, in the large $N$ expansion, the bare coupling constant
behaves as it would in an asymptotically free theory.  If we hold the
finite mass scale $\mu$ fixed as we put the ultraviolet cutoff
$\Lambda$ to infinity the bare coupling $\lambda$ runs to zero.  In
this sense, in the large $N$ expansion, the three dimensional
Gross-Neveu model behaves as if it were an asymptotically free theory.

\subsubsection{Massive phase with broken symmetry}

Now, we consider the regime where $\lambda>16\pi/\Lambda$ where
the gap equation for $M$ has a nonzero solution.  The correlator
$\tau(p)$ for this case can be computed.  It is
\begin{equation}
\tau_{\rm c}(q)=
- \frac{1}{16\pi} \Lambda e^{-\frac{2M}{\Lambda}} + \frac{1}{4\pi}
\frac{4M^2 + q^2}{|q|} \arctan\frac{|q|}{2M}
\label{t0}
\end{equation}

We combine this with the gap equation to find the effective coupling
constant
\begin{equation}
\lambda^{\rm eff}_{\rm c}(q) = \frac{4\pi}{
\frac{4M^2 + q^2}{|q|} \arctan\frac{|q|}{2M} }
\end{equation}
This coupling constant goes to zero like $\lambda^{\rm eff}_{\rm
c}\sim 16/|q|$ for large q and has a perfectly finite constant limit for
small q, $\lambda^{\rm eff}_{\rm c}\approx 4\pi/M$.  We note that both
the cutoff and the bare coupling constant have disappeared from this
equation.  It depends only on the fermion mass $M$.  This is the
analog of dimensional transmutation in this three dimensional model.

\subsection{Non-commutative theory}

In the non-commutative theory, there are two contributions to the
correlator $\tau(q)$, one from a planar diagram, which gives the same
result as in the commutative theory with an over all factor of 1/2,
\begin{equation}
\tau^{\rm pl}_{nc}(q)=
- \frac{1}{32\pi} \Lambda e^{-\frac{2M}{\Lambda}} + \frac{1}{8\pi}
\frac{4M^2 + q^2}{|q|} \arcsin\frac{|q|}{\sqrt{4M^2 + q^2}}
\label{t1}
\end{equation}
The other is from a non-planar diagram and depends explicitly on the
non-commutativity parameter,
\begin{eqnarray}
\tau^{\rm nonpl}_{\rm nc}(q)=
- \frac{1}{32\pi} \left(\frac{|\theta q|^2}{4}
+ \frac{1}{\Lambda^2}\right)^{-\frac12}
\exp\left[-2M \left(\frac{|\theta q|^2}{4}
+ \frac{1}{\Lambda^2}\right)^{\frac12}\right] + \nonumber \\
+ \frac{1}{16\pi} \left(4M^2 +
q^2\right) \int^1_0 \frac{d\alpha}{\sqrt{M^2 + \alpha(1 - \alpha) q^2}}
\exp\left(- |\theta q|\sqrt{M^2 + \alpha(1 - \alpha) q^2}\right)
\label{t2}
\end{eqnarray}
If we assume that the gap equation is satisfied, this theory also has
two different phases, which we can study separately.

\subsubsection{Symmetric, massless phase}

As in the commutative theory, if $\lambda<16\pi/\Lambda$,
the fermion mass is zero, $M=0$.
In this case we can use the massless limits of the equations (\ref{t1})
and (\ref{t2}) which are
\begin{equation}
\tau^{\rm nonpl}_{\rm nc}(q)=
- \frac{1}{32\pi} \Lambda + \frac{|q|}{16}
\label{t11}
\end{equation}
and
\begin{equation}
\tau^{\rm nonpl}_{\rm nc}(q)=
\frac{-1}{32\pi} \left(\frac{|\theta q|^2}{4} +
\frac{1}{\Lambda^2}\right)^{-\frac12}
+ \frac{|q|}{16\pi} \int^1_0 \frac{d\alpha}{\sqrt{\alpha(1 - \alpha)}}
\exp\left(- |\theta q| |q| \sqrt{\alpha(1 - \alpha)}\right) \label{t3}
\end{equation}
Then, the effective interaction is given by the equation
\begin{eqnarray}
\frac{1}{\lambda^{\rm eff}(q)} =
\frac{1}{{\lambda}}-\frac{\Lambda}{32\pi}
+ \frac{|q|}{16} - \frac{1}{32\pi}\left( \frac{|\theta q|^2}{4}+
\frac{1}{\Lambda^2}\right)^{-\frac{1}{2}}+
\nonumber \\ +
\frac{|q|}{16\pi} \int^1_0 \frac{d\alpha}{\sqrt{\alpha(1 - \alpha)}}
\exp\left(- |\theta q| |q| \sqrt{\alpha(1 - \alpha)}\right).
\label{end}
\end{eqnarray}
where we have kept the cutoff dependence in the third term on the
right-hand-side to regulate the behavior of this term at small $q$.
Now the question is, can we choose a cut-off dependent $1/\lambda$
in such a way that the large $\Lambda$ limit can be taken?

The answer is no.  The reason is because the right-hand-side of
(\ref{end}) has a minimum when $q=0$, so the smallest that we can make
it is to choose $1/\lambda$ to cancel its value at $q=0$.  This choice
is
$$
\frac{1}{\lambda}= \frac{\Lambda}{16\pi}+\mu^2
$$
Here, $\mu^2$ has to be positive if we are to be in the massless
phase, then $\mu^2\to0$ the coupling constant is at the critical
point.
\begin{eqnarray}
\frac{1}{\lambda^{\rm eff}_{\rm c}(q)} =
\mu^2
+ \frac{|q|}{16} +
\frac{1}{32\pi}
\left(
\Lambda-
\frac {1} {\sqrt { \frac{|\theta q|^2}{4}+
\frac{1}{\Lambda^2} } }
\right)^{-\frac{1}{2}}+
\nonumber \\ +
\frac{|q|}{16\pi} \int^1_0 \frac{d\alpha}{\sqrt{\alpha(1 - \alpha)}}
\exp\left(- |\theta q| |q| \sqrt{\alpha(1 - \alpha)}\right).
\label{end1}
\end{eqnarray}
This effective coupling is positive for all values of $q$.  This is
desirable, as this means that it does not have a Landau pole
singularity.

However, the cutoff dependence is still there.  For any non-zero value
of $q$ it behaves like
$$
\lambda^{\rm eff}(q)\approx \frac{32\pi}{\Lambda}
$$
which goes to zero as $\Lambda\to\infty$.  The theory has a
trivial effective coupling constant.

There is no freedom to further adjust $\mu^2$ so that the cutoff
dependence cancels since, to be in the massless phase, it must be
positive.  If we did, naively try to choose $1/\lambda$ so as to
cancel all of the cutoff dependence when the momentum is finite.  $
\frac{1}{\lambda}=\frac{\Lambda}{32\pi}+\mu^2 $ we would discover that
the massless ground state is unstable, that modes with momentum in the
range $\vert\theta q\vert\approx 1/16\pi\mu^2$ are tachyonic.

\subsubsection{Double Scaling limit}

As in the two dimensional case, there is a limit where the
cutoff can be decoupled.  Assume that $\theta^{0i}=0$ and $\theta^{ij}=
\epsilon^{ij}\theta$ where $\epsilon^{ij}$ is the fully antisymmetric tensor
with $\epsilon^{01}=1$.  We consider the limit where we take $\Lambda\to
\infty$ and then we keep $\Lambda^3\theta^2$ a fixed finite
number with momentum dimension -1.  
In that case,
$$
\frac{1}{\lambda_{\rm eff}(q)}= \frac{1}{\tilde\lambda}
+\frac{q}{8}+\frac{ \Lambda^3\theta^2 \vec q^{~2} }{256\pi}
$$
This means that the force between the fermions is mediated by
exchange of a particle with non-relativistic dispersion relation
$$
\omega(\vec q)=\vert\vec q\vert\sqrt{ \vec q^{~2}+\Lambda^3\theta^2/32\pi}
$$
It is long-ranged, but no longer scale invariant.
In this case, even though the non-commutativity occurs only at distance
scales that have been excised by the ultraviolet cutoff, there is still
some remnant of it caused by the ultraviolet divergences of the field
theory.  

This limit has rather severe implications when we consider the
space-space uncertainty relation $\Delta x\Delta y=\theta$ which we
re-write as $\Delta x\Delta y=(\Lambda^{3/2}\theta)/\Lambda^{3/2}$.
If $\Delta x$ is finite, then, there is a coordinate-momentum
uncertainty relation $\Delta x\Delta p_x\sim 1$ $\Delta y \sim
(\Lambda^{3/2}\theta) \Delta p_x/\Lambda^{3/2}$ is smaller than the
short distance cutoff which is of size $1/\Lambda$. The uncertainty
relation should therefore be invisible in the cutoff theory.  Note
that in the two dimensional theory it would have been of order the
small distance cutoff and marginally observable.

\subsubsection{Gapped phase with broken symmetry}

The massive phase of this theory suffers much the same fate as the
two-dimensional model.  The cutoff dependence of the bare coupling
constant is dictated by the gap equation.  We must solve the gap
equation to get a stable solution.  Then some cutoff dependence
remains in the effective coupling constant, From (\ref{t1}),(\ref{t2})
one immediately sees that for momenta above $|\theta q|\sim 1/\Lambda$
the dependence on momentum in $\lambda^{\rm eff}(q)$ is a small
correction in comparison with the term $\sim \Lambda$. Hence, we have

\begin{equation}
\lambda^{\rm eff}_{\rm c}(q) \approx
\frac{1}{\frac{1}{\lambda} - \frac{1}{32\pi}\Lambda e^{-\frac{2M}{\Lambda}}}
= \frac{32\pi}{\Lambda} e^{\frac{2M}{\Lambda}},
\end{equation}
where $\lambda$ is eliminated using (\ref{gap2}). Hence, in this phase
we get a situation similar to the 2-dimensional Gross-Neveu model,
except that the cutoff dependence is more severe - it effective
coupling goes to zero linearly, rather than logarithmically as the
cutoff is put to infinity.

\noindent
\section{Conclusions}

We have examined the issue of renormalizability of the large $N$
expansion of the non-commutative $O(N)$ Gross-Neveu models in 2 and 3
space-time dimensions.  In both cases, we find that the ultraviolet
cutoff does not decouple from the theory.  If we take it to infinity,
the four-fermion interaction becomes trivial.

There is a question as to whether this is a generic feature of
non-commutative field theories with translationary invariant vacua.
Of course, UV/IR mixing is generic.  Consider the following general
physical arguments.  An excitation with large momentum $p_x$ in a
non-commutative theory has uncertainty in its position along the
momentum $\Delta x \sim 1/p_x$ which is very small. Hence, taking into
account that $x$ and $y$ coordinates do not commute, we see that the
uncertainty in $y$ is very big. This mixes IR and UV limits in the
sense that IR effects modify UV limit and vice-versa \cite{Matusis:2000jf}.
This mixing is seen in our example as well as others as the appearance
of small momentum singularities in primitively divergent correlation
functions which are therefore also generic.

Generally, and especially in systems where the effective coupling is
strong in the infrared, the strong coupling dynamics generates a mass
gap.  In a commutative theory this generation of a mass gap is what
cures the problems, such as the infrared Landau pole, which occurs in
theories where the coupling constant has no infrared stable fixed
point.  In this Paper we have seen an explicit example of a
non-commutative theory where the same mechanism - the generation of a
mass gap - does not cure the problems associated with strong coupling.

Non-commutative gauge theories are often given as examples of field
theories which come from string theory in a strong field and zero
slope limit.  For example, in \cite{Gomis:2000bn} it was claimed that
one can obtain asymptotically free non-commutative SUSY Yang-Mills
theories by a self-consistent truncation of the massive string
modes. In this paper they extended tree level considerations of
\cite{Seiberg:1999vs} to the case of loop corrections.  It would be
interesting to check whether a consistent non-commutative theory
really results.  This is a very non-perturbative question 
~\cite{Khoze:2001sy} and is
complicated by the lack of any local gauge invariant observables in
non-commutative Yang-Mills theory \cite{Gross:2000ba} (see 
also \cite{Berenstein:2001dw}).
 
It is worth mentioning that we are aware of non-commutative theories
which are renormalizable because $\theta$ does not enter into
divergent graphs, i.e. non-commutativity does not modify UV
properties of the theories. The list of such theories includes
Gross-Neveu model with Dirac fermions and non-commutative
$N=4$ SYM theory living on the $D3$-brane in type IIB string theory
with constant $B$-field background.  In these cases, because the
ultraviolet divergences are unaffected by non-commutativity, the
infrared singularities that would appear with them are also absent.
 
An interesting aspect of our considerations for asymptotically
free non-commutative theories with $\theta$ parameter entering
into divergent diagrams is the assumption that the vacuum is translationally
invariant.  It indeed seems to be a reasonable assumption since the
effective action has a deep minimum for translationally invariant
states. 

After this paper was written, it was pointed out to us that there is
some overlap between our work and that contained in \cite{Girotti:2001gs}.
In fact, they confirm our analysis  of the Gross-Neveu model and they
also show that the 2-dimensional asymptotically free sigma models 
also have non-renormalizable large N expansions.

This work is supported in part by NSERC of Canada.  E.T.A. was
supported by a NATO Science Fellowship and grants INTAS-97-01-03 and
RFBR 98-02-16575 and G.S and E.T.A were supported in part by
NATO Collaborative Linkage Grant SA(PST.CLG.977361)5941. 
We thank Kostya Zarembo and Yurii Makkenko for valuable discussions.



\end{document}